# Surface Acoustic Wave induced modulation of tunneling magnetoresistance in magnetic tunnel junctions

Dhritiman Bhattacharya[1*], Peng Sheng[2*], Md Ahsanul Abeed[3], Zhengyang Zhao[2], Hongshi Li[2], Jian-Ping Wang[2*], Supriyo Bandyopadhyay[3*] Bin Ma[2*] and Jayasimha Atulasimha[1, 3*]

[1]Deparrtment of Mechanical and Nuclear Engineering, Virginia Commonwealth University, Richmond, VA-23284, USA
[2]Department of Electrical and Computer Engineering, University of Minnesota, Minneapolis, MN-55455, USA
[3]Department of Electrical and Computer Engineering, Virginia Commonwealth University, Richmond, VA-23284, USA
*Corresponding Author

We show that a surface acoustic wave (SAW) applied across the terminals of a magnetic tunnel junction (MTJ) decreases both the (time-averaged) parallel and antiparallel resistances of the MTJ, with the latter decreasing much more than the former. This results in a decrease of the tunneling magnetoresistance (TMR) ratio. The coercivities of the free and fixed layer of the MTJ, however, are not affected significantly, suggesting that the SAW does not cause large-angle magnetization rotation in the magnetic layers through the inverse magnetostriction (Villari) effect at the power levels used. This study sheds light on the dynamical behavior of an MTJ under periodic compressive and tensile strain.

**Introduction:**

Magnetic tunnel junction (MTJ) switches are *non-volatile*, which makes them attractive for many applications, but the non-volatility usually comes at an excessive energy cost (~10 – 100 fJ of switching energy) when the MTJ is switched with current controlled mechanisms like spin-transfer- or spin-orbit-torque [1-3]. Voltage-controlled switching modalities, on the other hand, are more energy-efficient [4-8] and among them straintronic switching [9-17] is exceptionally efficient in its use of energy. Consider an MTJ fabricated on a piezoelectric substrate with its (magnetostrictive) free layer in contact with the substrate as shown in Fig. 1(a). When a dc voltage is applied across the piezoelectric (not shown in Fig. 1(a)), a strain is generated in the substrate, which is transferred to the free layer and rotates its magnetization. This results in a change of the MTJ resistance and is the basis of straintronic switching. It is extremely energy-efficient with theoretically predicted energy dissipation of ~1 aJ per bit at room temperature [10].

Switching the resistance of an MTJ with voltage generated *static* strain has already been demonstrated [18, 19], but there is little experimental work that explores switching with *time-varying* strain, such as the one generated with a surface acoustic wave (SAW). In the past, SAW induced magnetization dynamics in Co microbars [20] and SAW based switching from single domain to vortex state [21] as well as complete magnetization reversal in isolated and dipole coupled Co nanomagnets were reported [22]. Furthermore, SAW-induced magnetization rotation and domain-wall motion in magnetostrictive nanomagnets have been harnessed for energy-efficient hybrid writing schemes for non-volatile memory [20]. Using standing or focused SAW, magnetic patterns were written on a film [23]. SAW based control of magnetization was also demonstrated in dilute semiconductors such as GaMnAs [24-25]. SAW of few MHz frequency can quasi-statically reduce the energy barrier within a nanomagnet to assist spin transfer torque-based switching [26] while SAW of few GHz can be used to excite acoustically driven ferromagnetic resonance in nanomagnets [27-29]. SAW of ~100 MHz frequency has been used to actuate extreme sub-wavelength magneto-elastic antennas consisting of magnetostrictive nanomagnets delineated on a piezoelectric substrate [30]. Optically generated SAW of several GHz frequency has been utilized to cause magnetization precession and generation of confined spin waves in isolated magnetic nanodots [31] as well as traveling spin waves in arrays of magnetic nanodots [32]. SAW-mediated picosecond magnetization switching has also been predicted [33-35].

In this study, we investigated the effect of SAW on the behavior of the resistance of MTJs by fabricating CoFeB/MgO/CoFeB based MTJs on a piezoelectric $LiNbO_3$ substrate. SAW was launched on the substrate using interdigitated transducers (IDTs). Although we did not observe any significant change in the coercivity of the magnetic layers due to SAW excitation, the MTJ resistance in both parallel and anti-parallel configurations and the tunneling magnetoresistance (TMR) ratio decreased. We attribute this effect to two causes: (1) the SAW imposes a time-varying bias voltage across the MTJ which reduces the time averaged resistance of both the parallel and anti-parallel configurations, despite the time-averaged value of the bias voltage being zero, because the effect exists for both polarities of the bias voltage [36, 37], and (2) time varying strain increases the time-averaged tunneling currents in both parallel and antiparallel spin channels, with the latter increasing more than the former [37]. As a result, the MTJ resistance decreases in both parallel and anti-parallel configurations, as does the TMR, when the MTJ is strained. The former effect is of electrical origin and the latter effect is of mechanical origin.

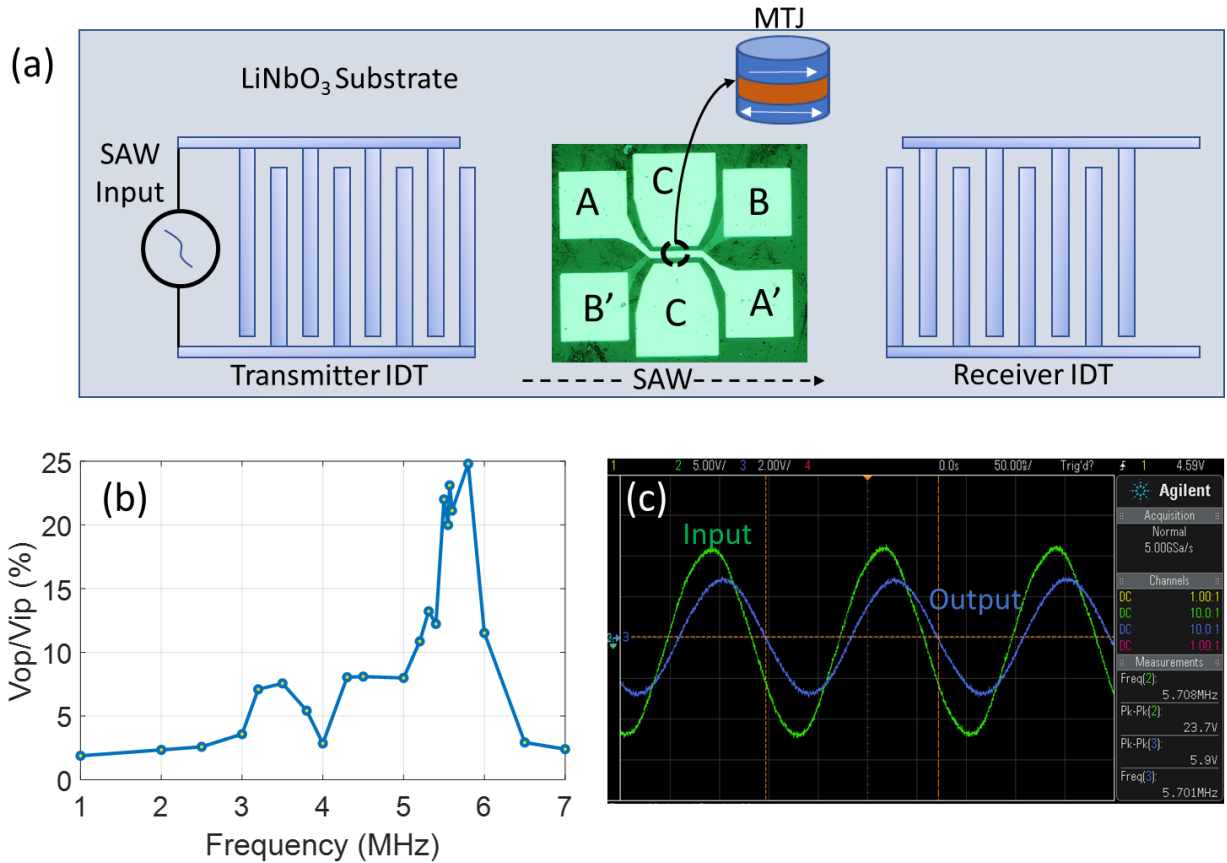

*Figure 1. (a)Schematic of the fabricated device. An array of MTJs is placed between a pair of interdigitated electrodes. Either A-B or B'-A' electrode pair was used to measure the MTJ resistance and they are in contact with the piezoelectric substrate. Electrode C was not used in this experiment, (b) The ratio of the output voltage to the input voltage (fixed) detected at the receiver IDT as a function of the input signal frequency. SAW resonant frequency is defined as the frequency where this ratio peaks. (c) Oscilloscope image of input and output signal waveforms (the scales are different for the two waveforms).*

Fig. 1(a) shows a schematic of the experimental set up. The interdigitated transducers (IDTs) are a comb like arrangement fabricated on a 128° Y cut LiNbO$_3$ substrate to generate SAW, using conventional photolithography and ion milling processes. We designed the characteristic frequency to be 3.7 MHz. The SAW propagation speed of ~3675 m/s would make this frequency corresponds to a wavelength of 1 mm. The pitch between the IDT fingers needs to be one-half the value of the SAW wavelength or 500 μm in our case. A metallization ratio of 0.75 is used while fabricating the fingers (i. e. width of rectangular aluminum bars is 375 μm and the gap between successive fingers is 125 μm). A pair of such IDTs (transmitter and receiver) were fabricated as shown in Figure 1(a). The transmitter is used to launch the SAW that interacts with the MTJs while the receiver transducer is used to confirm the propagation of SAW by measuring output voltage. An array of MTJs is fabricated in the delay line between the transmitter and the receiver IDT. The structure of the MTJ stack is Ta (8)/ Co$_{20}$Fe$_{60}$B$_{20}$ (2)/ MgO (2.3)/ Co$_{20}$Fe$_{60}$B$_{20}$ (5)/ Ta (8) (all thicknesses are in nm) from bottom to top, grown on the LiNbO$_3$ substrate. The MTJ pillars are circular or elliptical in shape and their dimensions vary from (2 μm × 2 μm) to (10 μm × 4 μm).

To confirm the propagation of a SAW through the substrate, an ac voltage of peak-to-peak amplitude 25 V (maximum) and frequency varying from 1-7 MHz is applied at one IDT and the output is measured at the other IDT. Fig. 1(b) shows the ratio of the output to input voltage (input amplitude fixed at 25 V) as the frequency is swept from 1 to 7 MHz. The output peaks at around 5.8 MHz which conforms to the resonant frequency of the fabricated IDTs. This confirms that a SAW is indeed launched in the substrate and it propagates. However, the SAW is significantly attenuated by the time it reaches the output IDT from the input IDT because the maximum ratio of output voltage to input voltage is only 25% (at the resonant frequency). This large attenuation is probably due to the fact that the substrate is damaged during the ion milling process employed to fabricate the MTJs. Fig. 1(c) shows the input and output waveforms at the two IDTs near resonance (frequency = 5.7 MHz).

At remanence (B = 0), the MTJ's free and fixed layer magnetizations are found to be almost anti-parallel because of dipole interaction between the two layers. We measured the anti-parallel resistance $R_{AP}$ (at remanence) between the electrodes *A* and *B* shown in Fig. 1(a) in the presence of SAW using a constant current source delivering 7 μA of current and varied the frequency and amplitude of the excitation voltage applied at the transmitter IDT shown in Fig. 1(a). These resistance measurement results are shown in Fig. 2. We note that the resistance $R_{AP}$ decreases in the presence of the SAW and the decrease is maximum (~65% decrease) at a frequency of 5.2 MHz which is close to the IDT resonance frequency of 5.8 MHz where the SAW amplitude is largest. The slight difference between the two frequencies (5.8 MHz and 5.2 MHz) can be attributed to the impedance loading effect of the measuring meter and the fact that the impedances of the IDTs and of the MTJ are different. We also varied the SAW excitation voltage amplitude and found that the resistance $R_{AP}$ decreases progressively with increasing amplitude. The resistance decreases by ~65% at the maximum applied amplitude of 25 V, with no sign of saturating at the maximum input SAW voltage used. These observations confirm that the SAW decreases $R_{AP}$ and the amount by which it decreases is maximum close to the resonant frequency (for a fixed input voltage). Also, the amount of decrease increases with input voltage amplitude (increasing SAW power) for a fixed frequency.

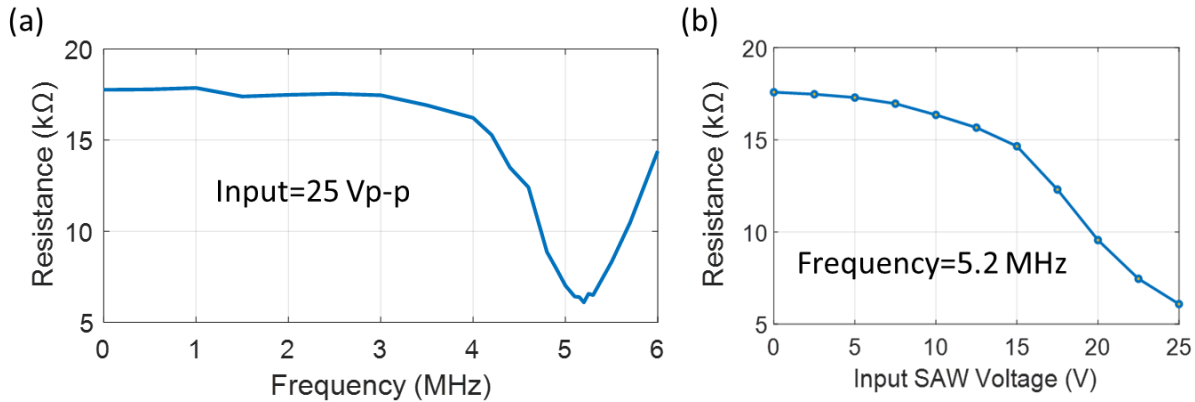

Figure 2. (a) Antiparallel MTJ resistance measured at remanence vs input SAW frequency. Resistance becomes minimum close to the SAW resonant frequency. (b) Antiparallel MTJ resistance vs input SAW voltage amplitude (peak-to-peak).

We then measure the tunneling magnetoresistance of MTJ with a magnetic field oriented along the easy axis. The MTJ resistance versus magnetic field plots are shown in Fig. 3(a). They are shown for two different scans (forward and backward) for two input voltages of 0 V and 10 V amplitude. There is a hysteresis between the scans, which is expected. Noting that the zero-field resistance is anti-parallel resistance ($R_{AP}$) and the high field resistance is the parallel resistance ($R_P$). In Fig. 3(b), we plot the TMR ratio as a function of the input SAW voltage. The TMR ratio, defined as the ratio $(R_{AP} - R_P)/R_P$, is 26% at 0 V (no SAW) and 3% at 25 V SAW excitation voltage.

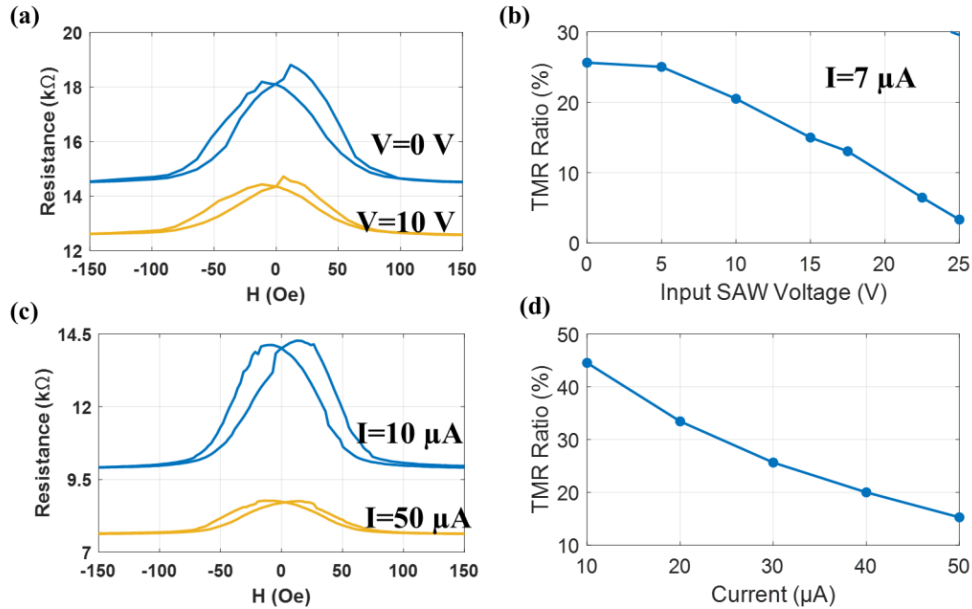

Figure 3. (a) Magnetoresistance curves at two different SAW voltages measured at 5.2 MHz, (b) TMR ratio vs input SAW voltage amplitude at 5.2 MHz, (c) Magnetoresistance curves at different bias voltages, (d) TMR ratio at different bias voltages

The first thing to note is that the switching of the MTJ configuration from anti-parallel to parallel by the magnetic field decreases the resistance by a maximum of 26%, which is *considerably less* than the decrease of 65% brought about by a SAW of 25 V amplitude and frequency 5.2 MHz at remanence. This tells us that SAW (of 25 V peak-to-peak amplitude) is *not* changing the MTJ resistance by rotating the magnetization of the free layer (because that would cause no more than 26% change), but because of some other effect. Note also from Fig 3(a) that the coercivities of the layers did not change perceptibly in the presence of the SAW.

The magnetoresistance curves in Fig. 3(a) are not sharp which indicates the presence of several intermediate stages in the magnetization reversal process. These intermediate stages could be switching of different magnetic domains. Although there are some reports on domain wall control due to SAW [20], this does not seem to be case here as the coercivity was unaffected. At different magnetic fields, we applied SAW and measured MTJ resistance before and after application of SAW. However, the resistance remained unchanged when there was no SAW applied. This means that the SAW was not able to affect the magnetization permanently or significantly.

The effect of the SAW is twofold. First, it applies an additional (time varying) voltage between the free and fixed layers of the MTJ in our device configuration (since the electrodes *A* and *B* rest on the substrate through which a SAW is propagating) and such a voltage was shown in the past to decrease both $R_{AP}$ and $R_P$, but the former is decreased much more than the latter [36, 37]. The decrease occurs for both polarities of the voltage and hence the effects during the positive and negative cycles of the time varying bias voltage do not cancel each other out. This will also decrease TMR. Second, strain (tensile or compressive) increases the tunneling probabilities of both parallel and anti-parallel spin channels, the latter more than the former [37] and this too causes the observed effects.

Bias voltage dependence of the magnitude and sign of the TMR has been studied in the ferromagnet/oxide interface [38-41]. Increased contributions from magnons, magnetic impurities, localized trap states, and the modified electronic structure at elevated biases and temperatures are responsible for the modulation of TMR with applied bias voltage [42]. This was also observed in our MTJs which is shown in Fig. 3(c, d). Note, this is a different MTJ from the one used in Fig. 3(a, b). Here, different currents were used to measure the TMR without launching any SAW. With increased current (i.e. increased applied bias voltage), both the TMR and the MTJ resistance $R_{MTJ}$ decreased similarly to what was observed in the case of SAW. For comparison, at I=50 µA, the change in TMR was three times while it decreased by eight times when input SAW signal was 25 $V_{P-P}$ (Fig. 3(d)). This lends credence to the first cause that we posited earlier, namely SAW generating an additional (time varying) voltage between the free and fixed layers of the MTJ decreases both $R_{AP}$ and $R_P$, but the former is decreased much more than the latter, so that the TMR ratio is decreased as well. The second cause, namely strain increasing the tunneling currents in both spin channels, with the current increasing more in the antiparallel channel, also probably exists but cannot be probed in our experiments.

An additional effect can impact our observations. The input IDT generates not only a SAW but also radiates an electromagnetic wave upon the application of the SAW excitation voltage since the IDTs can act as antennas. Electromagnetic pick-up between the MTJ electrodes *A* and *B* can contribute significantly to the measured output voltage in some circumstances. We discuss results in the last paragraph in the supplement to show that the SAW induced voltage plays a more significant role than the electromagnetic pickup.

In conclusion, this work demonstrates that a SAW can significantly change $R_{MTJ}$ and TMR. The effect of course has no memory since the original resistances and TMR are restored once the SAW is removed. In other words, the effect is volatile.

**Acknowledgement**: This work at VCU is supported by the US National Science Foundation under grant CCF - 1815033. The work at UMN is supported by the US National Science Foundation under grant CCF – 816406. Portions of this work were conducted in the Minnesota Nano Center, which is supported by the National Science Foundation through the National Nanotechnology Coordinated Infrastructure (NNCI) under Award Number ECCS-2025124.

# Supplementary Information

# Surface Acoustic Wave induced modulation of tunneling magnetoresistance in magnetic tunnel junctions

Dhritiman Bhattacharya[1*], Peng Sheng[2*], Md Ahsanul Abeed[3], Zhengyang Zhao[2], Hongshi Li[2], Jian-Ping Wang[2*], Supriyo Bandyopadhyay[3*] Bin Ma[2*] and Jayasimha Atulasimha[1, 3*]

[1]Deparrtment of Mechanical and Nuclear Engineering, Virginia Commonwealth University, Richmond, VA-23284, USA
[2]Department of Electrical and Computer Engineering, University of Minnesota, Minneapolis, MN-55455, USA
[3]Department of Electrical and Computer Engineering, Virginia Commonwealth University, Richmond, VA-23284, USA
*Corresponding Author

**SAW Characterization:**

An acoustic wave was generated and propagated along the substrate by applying a sinusoidal voltage between the two terminals of the transmitter IDT using a Tektronix CFG 280 pulse generator. In Figure 1(b) of the main paper, we plot the ratio of the output to input voltage amplitude (for a fixed input amplitude) vs input frequency. The output voltage is maximum at 5.8 MHz which is somewhat different from the designed resonant frequency of 3.7 MHz. Some variations in the experimentally observed resonance frequency from the designed one can be attributed to many factors including the SAW velocity being slightly different from the velocity assumed for a given crystallographic direction and deterioration of substrate properties during etching. The maximum output voltage amplitude is ~25% of the applied input voltage amplitude which indicates some loss of signal in the delay line. Attenuation of the acoustic wave depends on the frequency. At MHz frequency, significant attenuation of SAW signal is not expected. Etching of the substrate while fabricating the MTJs could have slightly deteriorated substrate properties which resulted in significant SAW attenuation in our case.

**Estimate of stress generated:**

The electrostatic potential generated in the piezoelectric substrate due to SAW propagation is $\phi = \mu_s(f,\eta) H(f)V$. Here $\mu_s(f, \eta)$ is the single-tap transmitter response function, $H(f)$ is the array factor, $\eta$ is the metallization ratio, $f$ is the frequency and $V$ is the amplitude of the applied sinusoidal voltage. When $f$ is equal to the resonant frequency of the SAW and a metallization ratio of 0.75 is used, $\mu_s(f,\eta) = 0.9K^2$, where $K^2 = 0.056$. The array factor $H(f)$ is equal to the number of pairs of electrodes in the transmitter IDT, which is 12 in our case. As 25% of the input voltage was transmitted, the resulting electrostatic potential associated with the SAW in the delay line of Lithium Niobate is 3.78 V when input voltage is 25 V. The particle displacement in the SAW is 0.18 nm per volt of electrostatic potential. The displacement wave can be expressed as $(3.78 \times 0.18) \sin(2\pi x/\lambda)$ nm, where the wavelength $\lambda$ is 1000 μm. Therefore, total displacement is 0.0128 nm over a length of 3 μm. The maximum strain generated by this acoustic wave in the substrate over a length of 3 μm length is 4.2 ppm. Assuming that this maximum strain is completely transferred to the nanomagnet, the maximum stress generated in the MTJ's free layer is ~0.8 MPa assuming Young's modulus is 190 GPa. This amount of strain was insufficient to cause significant rotation of the magnetization via the Villari effect. We can view stress acting as an effective magnetic field $H_{eff}$ given by $\mu_0 M_s H_{eff} = (3/2)\gamma_s \sigma$ where $M_s$ is the saturation magnetization ($1 \times 10^6$ A/m), $\gamma_s$ is the saturation magnetostriction (33 ppm) and $\sigma$ is the stress. The maximum value of $H_{eff}$ is 32 A/m or 0.4 Oe. This is obviously too small to cause any significant rotation of the magnetization of the MTJ's free layer via the Villari effect, consistent with our observation.

**Resistance measurement between terminals *A* and *B*, and results**:

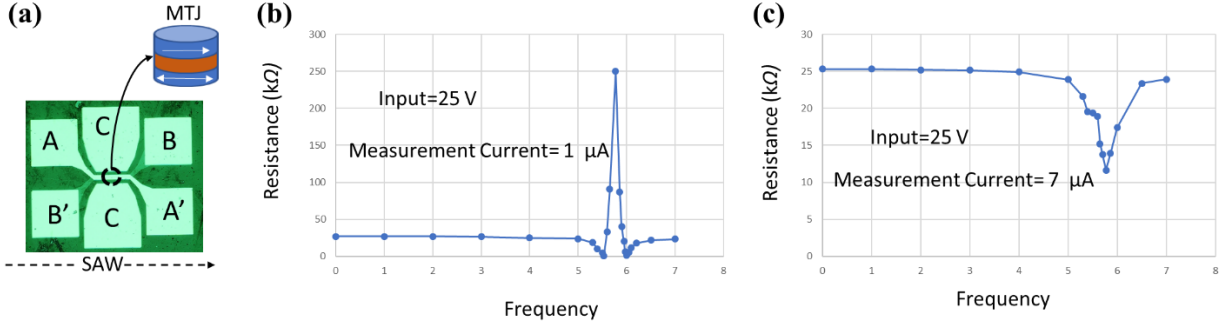

*Figure S1. (a) Schematic of a single MTJ with electrodes. (b) Time-averaged resistance between terminals A and B measured for one sample as a function of the SAW excitation frequency for a small constant current bias of 1 µA and SAW excitation amplitude of 25 V (c) Same measurement with a constant current bias of 7 µA.*

The resistance between terminals A and B is measured using a Keithley 2636 B electrometer and a Signatone probe station in a two-probe setup with a constant current.

In Fig. S1(b), we show the time-averaged resistance between terminals *A* and *B* measured for one sample as a function of the SAW excitation frequency for a small constant current bias of 1 µA and SAW excitation amplitude of 25 V peak-to-peak. In Fig. S1(c), we show results of the same measurement carried out at a constant current level of 7 µA. In the former case, the measured time-averaged resistance *peaks* at the IDT resonant frequency of 5.77 MHz while in the latter case, the resistance *troughs* at the frequency of 5.77 MHz. This dichotomy can be explained with the equivalent circuit shown in Fig. S2. There are two sources connected to terminals *A* and *B*: the voltage $V_{AB}^{SAW}(t)$ due to the SAW appearing between the MTJ terminals *A* and *B*, and the constant current source of the electrometer. The total voltage across the MTJ (i.e. between terminals *A* and *B*) is found by superposition:

$$V_{MTJ}(t) = V_{AB}^{SAW}(t)\frac{R_p||R_{MTJ}(t)}{R_p||R_{MTJ}(t)+R_s} + I\frac{R_s||R_p}{R_s||R_p+R_{MTJ}(t)}R_{MTJ}(t) \qquad (1)$$

where $R_{MTJ}(t)$ is the MTJ resistance, $R_S$ is the resistance associated with the electrodes and the material between the electrodes, $R_P$ is the internal resistance of the bias current source, $I$ is the constant current delivered by the current source and

$$V_{AB}^{SAW}(t) = V_0\bigl(sin(\omega t) - sin(\omega t + \theta)\bigr) \qquad (2)$$

where $V_0$ is the amplitude of the SAW induced voltage at any MTJ electrode and $\theta$ is the phase shift between the electrodes *A* and *B* of the MTJ. The time-averaged MTJ resistance measured by the dc meter is $R_{meas} = \frac{1}{T}\frac{1}{I}\int_0^T V_{MTJ}(t)dt$, where *T* is the period of the SAW. Note that in Equation (1), the sum is a phasor addition and there may be phase cancellation effects when the two terms are not in phase.

If the current source is near-ideal, then $R_P \to \infty$ and Equation (1) simplifies to

$$V_{MTJ}(t) = V_{AB}^{SAW}(t) + IR_{MTJ}(t) \qquad (3)$$

We know that $V_0$ and hence the amplitude of $V_{AB}^{SAW}(t)$ peaks at the resonant frequency while the amplitude of $R_{MTJ}(t)$ troughs at the resonant frequency. As a result, the amplitude of $V_{MTJ}(t)$ [and hence the measured time-averaged resistance $R_{meas}$] will be *highest* at the resonant frequency if the first term in Equation (3) dominates over the second term (which will happen at low values of the current *I*) and lowest at the resonant frequency if the second term dominates over the first term (which will happen at high current values). The former condition holds at the low current of 1 µA and the latter holds at the higher current of 7 µA. This explains the dichotomy observed in Figs. S1(b) and S1(c), i. e. the resistance peaks at the resonant frequency at low current levels and troughs at high current levels.

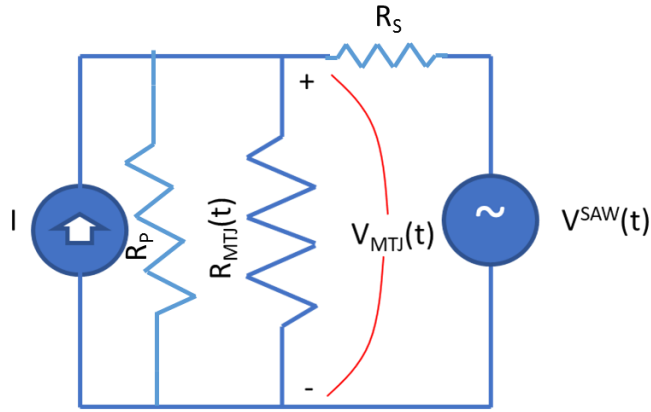

*Figure S2: Equivalent circuit to explain the observed results in Figs. S1(b) and S1(c).*

Note that only at *high current levels* the second term in Equation (3) dominates over the first, and only then is the time-averaged measured resistance $R_{meas} \approx \bar{R}_{MTJ}$, where $\bar{R}_{MTJ}$ is the actual time-averaged MTJ resistance. In the main text, we measured the resistance between terminals $A$ and $B$ at high current levels and hence we were measuring approximately the true MTJ resistance.

**Discussion of SAW induced voltage at the MTJ vs. Electromagnetic Pickup:**

It is hard to completely eliminate electromagnetic (EM) pickup in some circumstances, such as when the probes used to apply the voltage at the SAW transmitter act as a transmitting antenna that radiates an EM wave. The wave is picked up at the electrodes connected to the MTJ which act as a receiving antenna. However, in our case the SAW generated voltages dominate the EM pick up as can be confirmed by two observations:

1. Figure S1 (b) shows that the voltage picked up at the MTJ electrodes has a peak corresponding to the resonant condition of the SAW. If the pick-up voltage was dominated by EM pickup this could not be the case.

2. Figure S1 (b) shows a sinc function that is again typical of an IDT transfer function as a function of frequency and not EM waves. Hence the voltage between terminals A and B is affected much more by the SAW than any EM pickup.

3. No change in resistance was observed when input probes were held close to the input IDT but not touching it (i.e. no SAW was propagating). Change in resistance was only observed when input probes touched the IDT electrode. This proves that EM is not the dominating effect.

4. The IDT dimensions are much smaller than the EM wavelength at 5-6 MHz (which would be 50-60 meters). Hence the IDTs would be extreme sub-wavelength EM antennas with poor radiation efficiencies.